\documentclass[sigconf]{acmart}

\copyrightyear{2024}
\acmYear{2024}
\setcopyright{rightsretained}
\acmConference[Preprint Version, ESEM '24]{Proceedings of the 18th ACM / IEEE International Symposium on Empirical Software Engineering and Measurement}{October 24--25, 2024}{Barcelona, Spain}
\acmBooktitle{Proceedings of the 18th ACM / IEEE International Symposium on Empirical Software Engineering and Measurement (ESEM '24), October 24--25, 2024, Barcelona, Spain}\acmDOI{10.1145/3674805.3695405}
\acmISBN{979-8-4007-1047-6/24/10}

\setcopyright{none}

\usepackage{algorithmic}
\usepackage{graphicx}
\usepackage{xcolor,colortbl}
\usepackage{textcomp}
\usepackage{hyperref}
\usepackage{listings}
\usepackage{framed}
\usepackage{subfigure}
\usepackage{multirow}
\definecolor{codegreen}{rgb}{0,0.6,0}
\definecolor{codegray}{rgb}{0.5,0.5,0.5}
\definecolor{codeorange}{rgb}{1,0.49,0}
\definecolor{backcolour}{rgb}{0.95,0.95,0.96}

\usepackage{tikz}
\usepackage{pgfplots}
\pgfplotsset{compat=1.18}
\usepackage{mdframed}

\lstnewenvironment{logoutput}[1][]
{\lstset{basicstyle=\ttfamily, frame=lines, backgroundcolor=\color{gray!10}, #1}}
{}

\lstdefinestyle{mystyle}{
    backgroundcolor=\color{backcolour},   
    commentstyle=\color{codegray},
    keywordstyle=\color{codeorange},
    numberstyle=\tiny\color{codegray},
    stringstyle=\color{codegreen},
    basicstyle=\ttfamily\footnotesize,
    breakatwhitespace=false,         
    captionpos=b,                    
    keepspaces=false,                 
    numbers=none, 
    numbersep=5pt,                  
    showspaces=false,                
    showstringspaces=false,
    showtabs=false,                  
}

\AtBeginEnvironment{logoutput}{\footnotesize}

\usepackage{tcolorbox}

\newcommand{\boxit}[2][gray!15]{%
    \begin{tcolorbox}[
        colback=#1,    
        colframe=black, 
        width=8.3cm,    
        arc=0mm,        
        boxrule=0.5pt,  
        fontupper=\small,
        left=2pt,       
        right=2pt,      
    ]
        \emph{#2}
    \end{tcolorbox}
}

\definecolor{source}{gray}{0.9}

\begin{document}

\title{
From Struggle to Simplicity with a Usable and Secure API for Encryption in Java
}

\author{Ehsan Firouzi}
\affiliation{%
  \institution{Technische Universität Clausthal}
  \country{Germany}
}

\author{Ammar Mansuri}
\affiliation{%
  \institution{Technische Universität Clausthal}
  \country{Germany}
}

\author{Mohammad Ghafari}
\affiliation{%
  \institution{Technische Universität Clausthal}
  \country{Germany}
}

\author{Maziar Kaveh}
\affiliation{%
  \institution{Amazon Web Services}
  \country{USA}
}

 \lstset{language=Java,
    basicstyle=\sffamily\footnotesize,
	keywordstyle=\color{blue}\bfseries,
	mathescape=true,
	showstringspaces=false,
	keepspaces=true,
	numbers=left,                    
    numbersep=4pt,                  
	breaklines=true,
	breakautoindent=true,
	upquote=true, 
	columns=fullflexible} 
 \newcommand{\lct}[1]{{\textsf{\textup{#1}}}}
\lstnewenvironment{codesnippet}{%
	\lstset{%
		frame=single,
		framerule=0pt,
		mathescape=false
	}
}{}

\begin{abstract}

Cryptography misuses are prevalent in the wild.
Crypto APIs are hard to use for developers, and static analysis tools do not detect every misuse.
We developed SafEncrypt, an API that streamlines encryption tasks for Java developers.
It is built on top of the native Java Cryptography Architecture, and it shields developers from crypto complexities and erroneous low-level details. 
Experiments showed that SafEncrypt is suitable for developers with varying levels of experience.
 \end{abstract}

\ccsdesc[500]{Security and privacy~Cryptography}

\keywords{Cryptography, Encryption, Usable API}

\maketitle




\section{Introduction}

Cryptography APIs play a crucial role in the seamless integration of security into our digital world.
Nevertheless, the correct adoption of cryptography in software systems has proven to be challenging for developers.
Indeed, existing libraries often do not support auxiliary tasks, lack sufficient abstraction, and have poor documentation quality \cite{b5, mindermann2018}.  
Hence, mistakes in APIs, also referred to as ``API misuses'' in this paper, are likely and so does the presence of security vulnerabilities.
For instance, the analysis of 489 open-source Java projects has revealed that 85\% of cryptography APIs are misused~\cite{hazhirpasand2020}.
Similarly,
these misuses are common in online code examples, such as those found on the Stack Overflow website~\cite{firouzi2024time}.
Unfortunately, developers' mistakes in cryptography occur regardless of their experience~\cite{b18}.
There are static analysis tools aiming to uncover these misuses, but they have poor performance~\cite{b1,b2,b3}, and developers rarely adopt these tools~\cite{9680282}.


Symmetric encryption is one of the most frequently adopted cryptography operations in software systems.
%
%
%
We 
studied hurdles that developers face when adopting symmetric encryption in Java Cryptography Architecture (JCA) and investigated whether and how they are cleared in Google Tink, which is a modern encryption library.
We incorporated our findings into the design of a new API, called SafEncrypt, aiming to streamline the adoption of encryption for Java developers.
Through its step builder pattern, SafEncrypt offers task-based solutions and shields developers from
crypto complexities and erroneous low-level details.
In particular, it offers intuitive naming conventions, manages keys and IVs internally, resolves encoding and conversion issues, and provides informative exceptions.

To determine if SafEncrypt truly serves its purpose, we recruited ten developers and compared this API with JCA and Google Tink. The comparison was based on three different tasks, evaluating simplicity of use and time taken.
The results showed that SafEncrypt is beneficial for developers across all experience levels in developing symmetric encryption solutions. Remarkably, every participant found SafEncrypt easier to use than both JCA and Google Tink.

In summary, SafEncrypt, built on top of JCA, offers task-based solutions and hides details of low-level cryptography from developers. 
It is open source, and the API, its source code, documentation, and our analysis results are available on GitHub.\footnote{https://github.com/Ehsan-Firouzi/safencrypt}

The remainder of this paper is organized as follows.
We make an overview of related work in Section~\ref{sec:relatedwork}.
We investigate cryptography hurdles for developers in Section~\ref{sec:JCAChallenges}.
We present SafEncrypt in Section~\ref{sec:SafEncrypt} and explain its evaluation in Section~\ref{sec:Study}. 
We discuss threats to the validity of this study in Section~\ref{sec:ThreatstoValidity} and conclude this paper in Section~\ref{sec:Conclusion}.

\section{Related Work}
\label{sec:relatedwork}


We provide an overview of several factors that contribute to cryptography challenges for developers.

\emph{Lack of cryptography knowledge.}
Many developers using cryptographic APIs often lack important background knowledge, as shown by previous studies \cite{b4, b5}. 
However, developers should possess a solid background knowledge to use JCA effectively.

\emph{Documentation.}
Despite documentation being crucial for developers implementing libraries, challenges persist, as highlighted in studies \cite{b4, hazhirpasand2020, b5, b7}. Developers face difficulties with cryptographic operations due to insufficient documentation, and these challenges hinder the effectiveness of JCA implementation.

\emph{Usability.}
Several studies have explored usability issues in cryptography libraries. 
Nadi et al. aimed to investigate difficulties developers face in cryptographic implementation using JCA, analyzing posts and finding that developers, even with an understanding of cryptographic operations, made mistakes due to the complex APIs \cite{b4}. 
Green and Smith challenged the assumption that developers are inherently experts in cryptography, proposing ten guiding principles for designing secure and usable cryptographic APIs \cite{b10}. These principles emphasize ease of learning, self-explanatory design, error highlighting, prevention of misuse, and safe default settings.
Kafader and Ghafari introduced FluentCrypto, a modern API that eliminates many complexities~\cite{b11}. However, they built this API exclusively for Node.js program development.

\emph{Misuses and Challenges.}
Hazhirpasand et al.\cite{hazhirpasand2020} analyzed JCA misuse in 489 open-source projects on GitHub using CogniCrypt \cite{b13}. Common misuses included challenges with API parameters and insecure object passing. Their analysis found that only 15\% of repositories were free from misuse, leaving 85\% susceptible to security issues. In a specific study on GitHub projects, approximately 64\% of cryptographic solutions in each project were identified as not secure \cite{b18}. Additionally, Hazhirpasand et al. investigated cryptography-related questions on Stack Overflow (SO), revealing that the most common hurdles in implementing cryptographic solutions are the complexity of the underlying cryptographic API and developers' lack of familiarity with core cryptography concepts \cite{b12}.

\emph{Static Analysis Tools.}
Previous research showed that the adoption of static analysis tools is low among developers\cite{9680282}.
Indeed, these tools are not able to catch every mistake. 
Amit Seal Ami et al. highlighted a list of flaws where the detectors fail to detect vulnerabilities from a security perspective~\cite{b3}. 
Ying Zhang et al. analyzed and compared tools by running benchmarks and projects on several code scanning tools. Their subsequent user study highlights the concerns raised by developers on the reports generated by these tools\cite{b1}. Sharmin Afrose et al. developed detailed benchmarks and executed several vulnerability detection tools for the comparison of their effectiveness and reported their findings \cite{b2}. 
Overlooked issues include difficulties in resolving parameter values, insecure initialization vectors, insecure random number generation, and insufficient key lengths in cryptographic key generation.
Recently, researchers have proposed the adoption of Large Language Models (LLMs), such as ChatGPT, for program analysis~\cite{Kavian2024-lv}, and they explored their potential to detect cryptography misuse~\cite{firouzi2024time, Firouzi2024ChatGPT}. 
Their findings, based on relevant benchmarks, revealed that ChatGPT can be effective and even outperform state-of-the-art cryptography misuse detectors.

In summary, lack of background knowledge, poor documentation, usability issues, and limitations of static analysis tools have made encryption tasks challenging for developers.

\section{Developer Challenges}
\label{sec:JCAChallenges}

We aimed to attain a comprehensive understanding of developer challenges in symmetric encryption within the context of Java.
We investigated common issues that developers encountered when adopting JCA, the most popular and widely used cryptographic API in Java. 

We also investigated developer challenges with Google Tink, a modern library that claims to offer secure and easy to use cryptography.
We compared these two libraries to understand to what extent developer issues are cleared in Tink, what new challenges exist, and get inspired from its design.

\subsection{Methodology}
\label{sec:method}
We relied on the Stack Exchange
Data Dump released on March 8, 2023~\cite{ArchStackexchange} to uncover developer challenges in symmetric encryption discussed on the Stack Overflow website.

To identify relevant JCA posts, we searched the data dump for cipher instances instantiated with a symmetric encryption algorithm (shown in Listing~\ref{cipherpatternJCA}). 
We utilized the following regex to search within both questions and accepted answers. We discarded posts where the patterns were only found within other answers, as viewers of the posts pay greater attention to the aforementioned sections. 
This filtering process resulted in 3426 posts.
We chose a subset of these posts for a manual inspection.
In particular, we randomly included 400 posts (95\% confidence level with a less than 5\%(4.5\%) margin of error) to ensure that our findings represent the entire dataset, 
We sought to incline our sample data towards recent posts and those with high scores. Therefore,
we selected 40\% of the posts between 2020 and 2023, representing the most recent posts, 30\% from the most popular ones (highest scores), and the other 30\% completely random.

\begin{lstlisting}[label=cipherpatternJCA, caption=Regex for detecting symmetric encryption in JCA,language=C]
Cipher\.getInstance\(("|\&quot)(AES|DES|DESede|RC|Blowfish|ChaCha20)
\end{lstlisting}

%

To gather posts related to symmetric encryption in the Tink library, we searched the data dump for posts that included Tink as a tag. To collect posts that were relevant but did not include this tag, we searched the data dump for the symmetric encryption symptom shown in Listing~\ref{cipherpatternTink}. 
In the end, we obtained 73 relevant posts.
\begin{lstlisting}[label=cipherpatternTink, caption=Regex for detecting symmetric encryption in Tink,language=Java]
aead\.(encrypt|decrypt)\(
\end{lstlisting}

To categorize the root causes of JCA challenges and find out how Tink deals with these challenges, we conducted a lightweight open-coding-like process. This process involved four phases and was carried out by three individuals (i.e., A1, A2, and A3), each possessing practical knowledge in cryptography and over three years of Java programming experience. We describe the phases to conduct this qualitative study as follows:

\emph{Phase I.}
A1 and A2 collaboratively went through 50 questions and their associated accepted answers and comments. They collected developer challenges, guided by official sources \cite{javaReferenceGuide, ferguson2011cryptography}.
They also derived the intended encryption task from each post. 
Finally, they compiled two initial lists of challenges and intended encryption tasks.

\emph{Phase II.} A1 and A2 independently went through the remaining 350 posts and identified developer challenges as well as intended encryption tasks. 

\emph{Phase III.} A1 and A2  independently conducted a thorough review of all Tink posts, extracting insights into how Tink deals with challenges identified in Phase I.

\emph{Phase IV.} A1 and A2 compared their identified challenges and intended encryption tasks for JCA posts as well as insights obtained for each challenge from Tink posts. They discussed any disagreements until a consensus was reached. In cases where differing opinions persisted for certain posts, A3, who had not yet reviewed those specific posts, was consulted. Ultimately, the results were finalized using a majority voting mechanism (Cohen's k = 0.80).


\subsection{Results}

We present developer challenges in JCA and include any relevant observations related to Tink.


\begin{lstlisting}[label=lst:code1, caption=Key generation in Tink,language=Java, mathescape=true]
TinkConfig.register();
// 1. Generate the key material.
KeysetHandle keysetHandle = KeysetHandle.generateNew(
   AeadKeyTemplates.AES128_GCM);
// 2. Get the primitive.
Aead aead = keysetHandle.getPrimitive(Aead.class);
\end{lstlisting}

\subsubsection{Key Initialization}
\label{sec:KeyInitialization}

In\emph{JCA}, key generation involves low-level method calls, requiring an understanding of initialization and required key lengths.
We found that most
common root causes of developer problems resulted from key initialization (121 posts) such as invalid key sizes (length), password-based key derivation problems, and using different keys for encryption and decryption. 

\emph{Tink} simplifies key generation through its KeysetHandle primitive.
However, it seems to present a learning curve for its users.
Developers are not familiar with the naming conventions for Tink primitives. 
The use of AEAD (Authenticated Encryption and Associated Data) and the non-intuitive nature of Keyset and KeySetHandle concepts can confuse developers, especially those seeking simple symmetric encryption.
Listing~\ref{lst:code1} shows key generation in Tink.
First of all, there is a strict requirement to register the configuration.
Tink provides a KeysetHandle primitive in which the generate New method takes an algorithm from the KeyTemplate class as an input and returns a keyset. 
This keyset is then stored in a KeysetHandle holding the newly generated key by Tink.
Unlike JCA, Tink users do not encounter issues with key size as Tink handles it internally.
Tink challenges mostly revolved around differentiating between the concepts of keyset and key, as well as understanding how to generate keys with KeySetHandle, as highlighted in Stack Overflow posts [ID:72206958] and [ID:74395132]. 
In addition, Tink does not provide Password-Based Encryption (PBE) by default, as discussed in Stack Overflow posts [ID:52171198] and [ID:58972192].\\


\subsubsection{Initialization Vector (IV)}
\label{sec:IV}

We found that the second most prevalent issues in \emph{JCA} are related to IV (106 posts), such as how to generate IVs, the dependency of IVs on encryption modes, using different IVs for encryption and decryption, and considerations regarding the size of IVs.

\emph{Tink} automatically generates IV based on the specified algorithm during key generation, unless the user provides IV explicitly.
%
We did not find any questions regarding IV in Tink posts, which indicates that Tink performed well in this regard.

\subsubsection{Padding \& Encoding}
\label{sec:PaddingEncoding}

Padding (89 posts) and Encoding (54 posts) were two common challenging issues among \emph{JCA} users. 
The issues included choosing an improper padding mode, inconsistent padding modes for encryption and decryption, and employing different encoding systems for encryption and decryption.

\emph{Tink} handles padding internally, obviating the need for users to manage it. 
However, developers can mistakenly use different encoding systems for encryption and decryption, especially in interoperability scenarios. 
For example, in SO post ID 50693268, a user encountered challenges with a byte array to stream encoding issue while utilizing Google Tink. Similarly, in post ID 68593094, another developer confronted incorrect decryption output attributed to an encoding problem. 

\subsubsection{Encryption Mode.}
\label{sec:EncryptionMode}

We identified 49 \emph{JCA} posts that included problems with encryption mode such as the absence of an explicitly specified cipher mode, problems arising from dependencies between encryption mode and IV, and choosing different encryption modes for encryption and decryption.

\emph{Tink} 
does not require developers to specify encryption mode separately, minimizing the adoption of weak encryption configuration by mistake. 
Encryption modes such as ``AES\_GCM'', ``AES\_CTR'' or ``AES\_CFB'' are available to Tink users.
There was no Tink post about encryption mode.


\subsubsection{Key transmission}
\label{sec:KeyTransmission}

\emph{JCA} developers encountered challenges related to key transmission in 21 posts. 
They grapple with issues such as utilizing RSA and DH algorithms for exchanging symmetric keys and tackling keystore-related concerns.

 \emph{Tink} offers two methods for loading keysets, one for cleartext keysets and another for encrypted keysets. 
 The process involves using primitives like ClearTextKeySetHandle and AwsKmsClient. 
There was no direct question about key loading problems. However, questions arose due to confusion about the concept of keysets.

\subsection{Discussion}
   
\subsubsection{\emph{Security.}}

During our manual investigation of 400  \emph{JCA} posts, we observed that insecure practices are prevalent. Specifically, 34 posts utilized weak algorithms (DES, 3DES, RC1, RC2)(CWE-327), and 127 posts employed insecure ECB mode or failed to specify the encryption mode during instantiation (CWE-328).
It is noteworthy that if developers simply load the cipher instance 
Cipher.getInstance("AES") without indicating the mode of operation in JCA, they obtain the cipher instance in ECB mode. 
In addition, in 100 posts, static values were used for key initialization (CWE-798), 48 posts relied on static IV (CWE-330), and 33 posts utilized insecure algorithms (``PBKDF2WithHmacSHA1'' or ``PBEWithMD5AndDES'') for key generation in Password-Based Encryption (CWE-327).



\emph{Tink} prevents common security pitfalls such as the adoption of weak algorithms, using the ECB encryption mode, or using CBC encryption mode. 
Moreover, 
Tink mitigated risks related to keys and IVs through its design and the use of strong defaults.


\subsubsection{Interoperability.}
\label{sec:Interoperability}

We encountered 59 \emph{JCA} posts in which developers expressed concerns about interoperability.
There were 9 posts addressing issues with PHP, 9 posts related to Node.js, 8 posts highlighted challenges with C\#, and 33 posts were related to other programming languages.
We observed that the underlying causes of interoperability issues were primarily related to differences in default encoding and padding methods. 
We observed a consistent increase in concerns related to interoperability from 2020.

\emph{Tink} is designed to be cross-platform, with implementations available in multiple languages. This consistent API across different languages can be advantageous for projects that involve multiple platforms.
%
Nonetheless, interoperability between Tink and other libraries seems to be a non-trivial task due to its different design and conventions.
For example in SO Post IDs 71718371, 76656706, 74942579, and 76472518 the users expressed interoperability challenges with C\#. 
Tink is well-suited for closed environments like Google's backend, but it lacks widespread industry adoption.
The unconventional use of JSON format for its keys seems to be an important reason.


%

\subsubsection{Exceptions}
\label{sec:Exceptions}

We came across 128 exceptions during the inspection of \emph{JCA }posts (32\% of posts).
These exceptions are listed in Table~\ref{tab:SampExceptions}.
Our investigations revealed that the root cause for an exception may differ from the exception message, which can be confusing.
We found 61 cases where a \texttt{BadPaddingException} occurred. 
However, only 19 cases were directly related to padding problems. 
Most notably, 17 exceptions occurred during the encryption and decryption stages, primarily due to incorrect encoding.
We identified 14 cases with key-related issues, nine cases involving IV problems, and two other problems. 
In 45 instances where an \texttt{InvalidKeyException} occurred, only 26 exceptions were directly linked to key initialization problems. 
Eight exceptions were thrown in the context of IV issues, while the remaining 11 exceptions were associated with other problems.

 \begin{table}[htbp]
  \hspace{1cm}
  \caption {The top three recurrent exceptions}
 \begin{tabular}{|l|l|p{3.7cm}|} 
 \hline
  \textbf{Exception Type}                           &  \textbf{\#}                   &  \textbf{Exception Message (\#)}                          \\ \hline
 \multirow{2}{*}{BadPaddingException}       & \multirow{2}{*}{61} & Given final block not   properly padded (36)  \\ \cline{3-3} 
                                            &                     & pad block corrupted (13)                      \\ \hline
 \multirow{6}{*}{InvalidKeyException}       & \multirow{6}{*}{45}  & Illegal key size (14)                         \\ \cline{3-3} 
                                            &                     & Invalid AES key length (9)                    \\ \cline{3-3} 
                                            &                     & Key length not 128/192/256   bits (6)         \\ \cline{3-3} 
                                           &                     & Parameters missing (6)                        \\ \cline{3-3} 
                                            &                     & No installed provider supports   this key (3) \\ \cline{3-3} 
                                            &                     & Wrong Algorithm (3)                           \\ \hline
 \multirow{2}{*}{IllegalBlockSizeException} & \multirow{2}{*}{19} & Input length must be   multiple of 16 (10)    \\ \cline{3-3} 
                                            &                     & Last block incomplete in   decryption (3)     \\ \hline
 \end{tabular}

 \label{tab:SampExceptions}
 \end{table}

\emph{Tink} consolidates exceptions within its GeneralSecurityException class for a cleaner approach. Tink exceptions do not seem to be as confusing as JCA exceptions.
However, they do not pinpoint the exact problem, making it challenging for novices to troubleshoot the problem, Also  
For example, in Stack Overflow post ID=58680609, the ``java.security.GeneralSecurityException: decryption failed'' message does not reveal differences in encryption and decryption keys. 


\boxit[yellow!10]{
To sum up,
JCA remains a more popular and widely used library compared to Tink~\cite{CompareTrends}. However, JCA requires developers to manage low-level details, which can introduce security risks if not handled properly. In contrast, Tink offers a secure API that addresses many of these issues. Nonetheless, it may not be very user-friendly, particularly for novices. Users of Tink need a background in cryptography, which is often lacking among beginners. Additionally, Tink introduces unconventional primitives and a new key format, which can result in a steep learning curve. Tink also has limited resources, and its limitations in terms of readability, exception handling, and support have been noted in previous studies~\cite{cryptolib}.
}


\section{\NoCaseChange{SafEncrypt}}
\label{sec:SafEncrypt}

The challenges developers face when using JCA as a traditional library and Tink as a modern library inspired us to design 
\emph{SafEncrypt}, a wrapper built on top of JCA, aiming to shield developers from the intricacies of cryptography and the low-level implementation details that often lead to errors.
It enables developers to adopt secure encryption, even without background knowledge, and it is compatible with the widely used JCA, making its integration even into legacy projects seamless.
%
%
Instead of requiring developers to navigate through numerous steps and parameters, SafEncrypt employs a step builder pattern to streamline the encryption process into a single, cohesive API. This approach simplifies the user experience, providing clear guidance at every stage and minimizing the likelihood of mistakes.
With SafEncrypt, developers only need to specify their encryption requirements, letting the library handle the rest. By guiding users through operations in a step-by-step manner and narrowing down choices, SafEncrypt ensures clarity, reduces errors, and delivers a seamless user experience from start to end.
Table~\ref{tab:compare} shows a comparison of some features among SafEncrypt, Tink, and JCA.

\begin{table*}[]
\caption{Comparison of some technical features: JCA, Tink, and SafEncrypt}
\begin{tabular}{|l|l|l|l|}
\hline
{\color[HTML]{222222} \textbf{Feature}}         & \textbf{SafEncrypt}                                                                                      & \textbf{Tink}                                                                                        & \textbf{JCA}                                                                                                \\ \hline
{\color[HTML]{222222} \textbf{Lines of Code}}   & Fewest                                                                                                   & Moderate                                                                                             & Most                                                                                                        \\ \hline
{\color[HTML]{222222} \textbf{Key Generation}}  & {\color[HTML]{222222} \begin{tabular}[c]{@{}l@{}}Handled internally\\  via builder pattern\end{tabular}} & {\color[HTML]{222222} \begin{tabular}[c]{@{}l@{}}Simplified with \\ Tink key templates\end{tabular}} & {\color[HTML]{222222} Requires  explicit setup}                                                             \\ \hline
{\color[HTML]{222222} \textbf{Initialization}}  & {\color[HTML]{222222} \begin{tabular}[c]{@{}l@{}}Simplified with \\ fluent builder pattern\end{tabular}} & {\color[HTML]{222222} \begin{tabular}[c]{@{}l@{}}Requires setup \\ of Tink primitives\end{tabular}}  & {\color[HTML]{222222} \begin{tabular}[c]{@{}l@{}}Requires detailed setup\\  and configuration\end{tabular}} \\ \hline
{\color[HTML]{222222} \textbf{Method Chaining}} & {\color[HTML]{222222} Yes (fluent API)}                                                                  & \cellcolor[HTML]{FFFFFF}{\color[HTML]{222222} No (separate method calls)}                            & \cellcolor[HTML]{FFFFFF}{\color[HTML]{222222} No (separate method calls)}                                   \\ \hline
{\color[HTML]{222222} \textbf{Data Conversion}} & {\color[HTML]{222222} Internally   managed}                                                              & \cellcolor[HTML]{FFFFFF}{\color[HTML]{222222} Manual conversion}                                     & \cellcolor[HTML]{FFFFFF}{\color[HTML]{222222} Manual conversion}                                            \\ \hline
\end{tabular}
\label{tab:compare}
\end{table*}

SafEncrypt currently supports symmetric and streaming symmetric encryptions, including all variants of AES-GCM (Galois/Counter Mode) and AES-CBC (Cipher Block Chaining).
We provided comprehensive information regarding existing algorithms, default options, configuration process, common usage scenarios, and reflected on the applicability of SafEncrypt to address real-world scenarios on the project's FAQs.\footnote{https://github.com/Ehsan-Firouzi/safencrypt}

We describe the configuration of SafEncrypt and how we ensure its security and ease of use in Section \ref{sec:Configuration}
We delve into understanding and resolving exceptions in Section \ref{sec:SafEncExceptions} 
Finally, in Section \ref{sec:WorkingInterface}, we explain how developers can adopt SafEncrypt in action.

\subsection{Configuration}
\label{sec:Configuration}

SafEncrypt offers cryptographic operations that are currently considered secure, and they can be configured in the future if needed. 
It relies on a whitelisting approach to determine the adoption of trusted configurations such as secure algorithms. 

There is an Enum class corresponding to every configuration file, restricting developer options to only trusted configurations.
%
%
SafEncrypt only accepts parameters from the associated ENUM classes, and it ensures that any parameter listed in the ENUM class is in sync with the one listed in the configuration file.
Otherwise, it does not allow developers to continue the encryption process.
In general, the library maintainers are responsible for configurations. Nonetheless, if desired, developers can also update the configurations in the JSON files.

Depending on the IDE, available configurations might be shown to developers through a drop-down list, allowing them to easily select from available secure options during coding. Nonetheless, at runtime, we check whether valid configurations are adopted, which prevents developers from making mistakes (e.g. if they did not use an IDE that shows valid options) or attackers from injecting weak configurations.

\subsubsection{Algorithms}
\label{sec:Algorithms}
There is a specific format of the algorithms that are used by SafEncrypt to ensure readability. An example is “AES\_CBC\_128\_PKCS5Padding” which interprets as AES encryption in CBC Mode using PKCS5Padding with 128 bits of key.
It offers a range of secure algorithms with correct key size, mode of encryption, and padding.
Listing~\ref{lst:code19} shows a configuration file for symmetric encryption.
It contains a list of all the algorithms that are trusted. Moreover, there are other configurable attributes associated with the algorithms. These configurations are dynamic and can be modified directly for future upgrades to avoid mass changes within the library. As AES\_CBC uses an IV which is 16 bytes of length it is defined under the constraints of AES\_CBC. Constraints for AES\_GCM define the length of the IV as well as the TAG\_LENGTH that are required during the encryption and decryption phases. 


\begin{lstlisting}[float, label=lst:code19, caption=Configurations for symmetric algorithms,language=Java, mathescape=true]
{
  "symmetric-algorithms": [ "AES_CBC_128_PKCS5Padding",
  "AES_CBC_192_PKCS5Padding", "AES_CBC_256_PKCS5Padding",
  "AES_CBC_128_PKCS7Padding", "AES_CBC_192_PKCS7Padding", 
  "AES_CBC_256_PKCS7Padding", "AES_GCM_128_NoPadding",
  "AES_GCM_192_NoPadding", "AES_GCM_256_NoPadding"
  ],
  "constraints": {
    "AES_CBC": {
      "iv-bytes": 16
    },
    "AES_GCM": {
      "iv-bytes": 12,
      "tag-bits": 96
    }
  }
}
\end{lstlisting}

Algorithms added under the symmetric algorithms are trusted, and they are also required to be added in a Java ENUM class shown in Listing~\ref{lst:code18}.

\begin{lstlisting}[float, label=lst:code18, caption=ENUMs for symmetric algorithms,language=Java, mathescape=true]
public enum SymmetricAlgorithm {
    //Correct Algorithms Currently Supported and ENABLED to promote Interoperability
    AES_CBC_128_PKCS7Padding("AES_CBC_128_PKCS7Padding"),
    AES_CBC_192_PKCS7Padding("AES_CBC_192_PKCS7Padding"),
    AES_CBC_256_PKCS7Padding("AES_CBC_256_PKCS7Padding"),
    //Correct Algorithms Currently Supported and ENABLED
    AES_CBC_128_PKCS5Padding("AES_CBC_128_PKCS5Padding"),
    AES_CBC_192_PKCS5Padding("AES_CBC_192_PKCS5Padding"),
    AES_CBC_256_PKCS5Padding("AES_CBC_256_PKCS5Padding"),
    AES_GCM_128_NoPadding("AES_GCM_128_NoPadding"),
    AES_GCM_192_NoPadding("AES_GCM_192_NoPadding"),
    AES_GCM_256_NoPadding("AES_GCM_256_NoPadding"),
    DEFAULT("AES_GCM_128_NoPadding");
    }
\end{lstlisting}



\subsubsection{Common tasks}
\label{sec:CommonTasks}
 
SafEncrypt includes 
configurations for tasks such as key generation.
Listing~\ref{lst:pbe} shows the configuration for password-based key generation, and Listing~\ref{lst:code20} shows its corresponding enum class.
During a password-based key generation, it is critical to define the correct configuration for the salt length and the number of iterations required to ensure security. 
Likewise other parameters, these values can be configured if needed.

\begin{lstlisting}[float, label=lst:pbe, caption=Configurations for PBKDF2 algorithms,language=Java, mathescape=true]
{
  "algorithms": [
    "PBKDF2WithHmacSHA256",
    "PBKDF2WithHmacSHA512"
  ],
  "salt-bytes": 64,
  "iterations": 1024
}
\end{lstlisting}


\begin{lstlisting}[float, label=lst:code20, caption=Enums for PBKDF2 algorithms,language=Java, mathescape=true]
public enum KeyAlgorithm {
    PBKDF2_With_Hmac_SHA256("PBKDF2WithHmacSHA256"),
    PBKDF2_With_Hmac_SHA512("PBKDF2WithHmacSHA512"),
    DEFAULT("PBKDF2WithHmacSHA512");
  }
\end{lstlisting}


\subsubsection{Secure defaults}
\label{sec:SecureDefaults}
 SafEncrypt is designed to contain pre-configured default configurations that tend to remove the complexity of creating significant parameters from the developer's end such as salts, key, IV, etc., and constraints the developers in the algorithms offered from a range of secure algorithms. In SafEncrypt, the configured default algorithm unless the developers specify during the encryption/decryption process is “AES\_GCM\_128\_NoPadding” [AES in GCM Mode with No Padding using 128-bit key length], which is configurable as well. Moreover, SafEncrypt is capable of generating all of the required security parameters internally depending on the algorithm. 
 For the key, developers just have to specify whether they want the default symmetric key generation, or they want to provide a password if they intend to use password-based key derivation features. During encryption, there is no IV in question. Generating all these parameters internally ensures their secure generation by using appropriate secure random mitigating possible misuses. With the introduction of secure defaults,\emph{developers even without knowledge of cryptography can use SafEncrypt to encrypt/decrypt their data securely}.

\subsubsection{Interoperability}
\label{sec:SafEncInteroperability}

%
SafEncrypt supports customized configuration, as shown in listing ~\ref{lst:code21}, to enable interoperability.
Developers set the parameters according to the programming languages that they intend to use so that the operations they perform remain fully compatible with those languages.
Developers just have to define during the symmetric encryption/decryption phase if they want to make it interoperable with the available choice of programming languages defined in the configuration. When the developers decide to make their cryptographic operations interoperable with any other programming language providing SafEncrypt the language of their choice as a parameter, it directly loads the user-defined configuration for that language from the configuration file.

We did not overload interoperability in the main call because making the same call as ``SafEncrypt.symmetricEncryption()'' for interoperability would again present a lot of options for the users, potentially confusing. Instead, we opted for a separate call to make it easier and separate the functions responsibly.


 \subsection{Exceptions}
 \label{sec:SafEncExceptions}

We support developers in understanding the obscured reasons behind JCA exceptions (see Section~\ref{sec:Exceptions} for details).
We collected possible exceptions with their causes that are thrown by JCA for symmetric encryption and created concrete error messages for each of them attached with an exception code. SafEncrypt then internally maps the exceptions thrown by JCA to the corresponding exception in SafEncrypt. For exceptions that arise from JCA, SafEncrypt throws a customized exception code and message in a specific format depicted in Listing~\ref{log:log2}.
%
The exception is interpreted as [ JCA Exception: JCA Exception Message] | [SafEncrypt Exception : SafEncrypt Exception Message] which gives the messages both from JCA and SafEncrypt for clarity purposes. In addition to that SafEncrypt also provides compile time checking for multiple circumstances before the actual encryption/decryption phase using a fail-fast strategy. Listing~\ref{log:log3} portrays an exception with a concrete exception message for a scenario where a key with inadequate length is provided for symmetric decryption while the algorithm selected for decryption in SafEncrypt is “AES\_GCM\_256\_NoPadding”. All the probable exceptions from SafEncrypt are accumulated in a configurable file
with the error codes and their corresponding messages with the possibility of making them personalized. 

\begin{logoutput}[float, label={log:log2}, caption={Exception format in SafEncrypt
%\mg{we include the possible causes behind an exception. AFAIK one can't get the exact issue until going deep into debugging and finding which method is exactly throwing the exception}
},captionpos=b]
[javax.crypto.BadPaddingException: Given final block not properly padded. Such issues can arise if a bad key is used during decryption.] | [SAF_010 : Either the Mode/Key/IV/Padding used for encryption was different than provided for decryption]
\end{logoutput}

\begin{logoutput}[float, label={log:log3}, caption={Key length exception in SafEncrypt},captionpos=b]
[SAF_003 : Provided Key With Length [23] bytes is not compatible with selected algorithm [AES_GCM_256_NoPadding], it should be exact [32] bytes long]
\end{logoutput}


\begin{lstlisting}[float, label=lst:code21, caption=Part of interoperability configuration,language=Java, mathescape=true]
{
  "interoperable-languages": {
    "Python": {
      "library-Provider": "Crypto",
      "symmetric": {
        "default-algo": "AES_CBC_256_PKCS7Padding",
        "iv-bytes": 16
      }
    },
    "CSharp": {
      "library-Provider": "New Library",
      "symmetric": {
        "default-algo": "AES_GCM_256_NoPadding",
        "iv-bytes": 12,
        "tag-bits": 96
    ...
\end{lstlisting}

It is important to provide a trade-off between flexibility and security.
To accommodate such scenarios, configurations can be accompanied by a compilation warning.
For instance, 
``AES'' symmetric encryption in CBC mode is not secure only in client-server architecture.
If it is not whitelisted, it reduces the flexibility of the API. 
Thus, it is whitelisted with a warning, shown at compilation time (Listing~\ref{log:log1}).

\begin{logoutput}[label={log:log1}, caption={Warning for CBC mode in client/server scenarios},captionpos=b]
[main] WARN com.safencrypt.service.SymmetricImpl - [SAF_011 : Usage of Algorithm [AES/CBC] is insecure in client-server architecture]
\end{logoutput}

\subsection{Working Interface}
\label{sec:WorkingInterface}



We adopted method chaining, where each method serves a particular operation required for encryption/decryption. Thanks to modern IDEs like IntelliJ, these operations are shown to developers step by step during coding. If any operation is forgotten, there will be compilation errors, ensuring smooth and error-free implementation.

Listing~\ref{lst:code26} shows an example usage of SafEncrypt with its default configurations. 
When the user types ``SafEncrypt'', a series of available operations is shown to her. Selecting symmetricEncryption() in the first step initiates the default algorithm behavior.
When using modern IDEs, there is no need to memorize complex strings like ``AES\_CBC\_256\_PKCS7Padding''. For instance, once developers type in ``AES'', SafEncrypt provides a list of secure options to choose from.
In the second step, opting for generateKey() automatically generates a random key for symmetric encryption. The key length is determined by the algorithm specified in the symmetricEncryption() method's parameter, defaulting otherwise. The third step involves entering plain text using the plaintext() method. In the final step, the user selects encrypt(), prompting SafEncrypt to perform encryption and provide detailed decryption information.

Similarly, for decryption, the user starts by typing ``SafEncrypt'', revealing available operations. Choosing symmetricDecryption() in the first step triggers the default algorithm behavior. In the second and third steps, the user is guided to enter the key and IV. The fourth step involves entering the encrypted text using the cipherText() method. In the fifth step, the user selects decrypt(), prompting SafEncrypt to perform decryption and return the plain text, concluding the operation.


 \begin{lstlisting}[float, label=lst:code26, caption=Basic encryption/decryption example,language=Java, mathescape=true]
 byte[] plainText = "Hello World".getBytes(StandardCharsets.UTF_8);
        SymmetricCipher symmetricCipher =
                SafEncrypt.symmetricEncryption()
                        .generateKey()
                        .plaintext(plainText)
                        .encrypt();

        byte[] decryptedText =
                SafEncrypt.symmetricDecryption()
                        .key(symmetricCipher.key())
                        .iv(symmetricCipher.iv())
                        .cipherText(symmetricCipher.cipherText())
                        .decrypt();

\end{lstlisting}

\section{Study}
\label{sec:Study}

We investigated SafEncrypt's applicability in addressing real-world scenarios collected from StackOverflow and conducted a user study to compare SafEncrypt with JCA and Tink libraries.

\subsection{Encryption Tasks in the Wild} \label{sec:RealWorldScenarios}

We identified 43 intended tasks from 400 StackOverflow posts in section~\ref{sec:method}.
To investigate the applicability of SafEncrypt in implementing real-world encryption scenarios, we examined whether it can address these encryption tasks.

Two participants, A1 and A2, independently 
provided their solutions for the tasks, and in the end, they compared them together.
%
They did not encounter any disagreement. 
They could adopt SafEncrypt and provide a proper working example for every task that was secure.
However, there were no working examples for a small subset of tasks that requested not enough secure practices (e.g., ``How do I specify my Cipher object to use DESede?'').
Indeed, the configuration of SafEncrypt does not allow setting up a cipher object which is not secure. If required, users have the flexibility to adjust the configurations to meet their specific needs.

In summary, our findings indicate that SafEncrypt is capable of handling real-world encryption tasks. This outcome was expected, given that we developed SafEncrypt based on challenges that developers faced in real-world.
We have included the tasks along with hints regarding how to use SafEncrypt to address each task in the FAQ section of the README  file in our replication package.

\subsection{User Study}

In this study, participants were asked to perform one randomly assigned task from a set of three common tasks (Simple Text Encryption, Password-Based Encryption, File-Based Encryption) using three different libraries (JCA, Tink, and SafEncrypt). After completing the task, they were required to fill out a survey questionnaire.

\subsubsection{Methodology}
Our Methodology consists of the following steps:
\\
    1- We initially invited a number of Java developers to participate in our survey. Invitations were sent to developers from various companies with different levels of experience and expertise. Since our main aim was to create a simple-to-use cryptography API for novices, expertise in cryptography was not a selection criterion. Participants were instructed to save their code and record the time from the start to the completion of each task, following a specific format. The importance of recording this time was explained to them.\\ 
    2- We provided links to the documentation for JCA, Tink, and SafEncrypt, allowing participants to familiarize themselves with these libraries and prepare for the tasks.\\
    3- Each developer was randomly assigned one task from our set of three tasks and was asked to implement it using JCA, Tink, and our library (SafEncrypt).\\ 
    4- Upon completing the task, participants filled out a questionnaire. This questionnaire had two sections: one related to the participant's background and the other specifically about the tasks.\\ 
    5- In the background knowledge section, participants were asked to provide (1) \emph{their years of experience in Java and their level of expertise}, and (2) \emph{any prior experience with cryptographic APIs}.\\
    6- In the task-specific section, participants provided the following information for each task: \emph{the difficulty level on a linear scale from 1-5 (Very Easy to Very Difficult)}, \emph{factors influencing their difficulty rating (optional)}, \emph{the time taken to complete the task}, \emph{the code they wrote for the task}, and \emph{which library they found easier to use and preferred}.\\   
    7-Additionally, participants were asked if they encountered any specific challenges with SafEncrypt (optional) and if they had any comments or suggestions (optional).\\

\begin{table*}[]
\caption{Participants\\
(For JCA and Tink Cryptography 
knowledge 0 = No, 1= Somewhat and 2 = Completely)}
\begin{center}
\begin{tabular}{|cl|l|lllllllllll|}
\hline
\multicolumn{2}{|c|}{\multirow{2}{*}{\textbf{Java}}}         & \multicolumn{1}{c|}{\multirow{3}{*}{\textbf{\begin{tabular}[c]{@{}c@{}}Number of \\ Participants\end{tabular}}}} & \multicolumn{11}{c|}{\textbf{Cryptography  Knowledge}}                                                                                                                                                                                                                                                                                                                                                                                                                              \\ \cline{4-14} 
\multicolumn{2}{|c|}{}                                       & \multicolumn{1}{c|}{}                                                                                             & \multicolumn{5}{c|}{\textbf{Concepts}}                                                                                                                                                                                                   & \multicolumn{3}{c|}{\textbf{JCA}}                                                                                              & \multicolumn{3}{c|}{\textbf{Tink}}                                                                        \\ \cline{1-2} \cline{4-14} 
\multicolumn{1}{|l|}{\textbf{Experience}}   & \textbf{Level} & \multicolumn{1}{c|}{}                                                                                             & \multicolumn{1}{l|}{\textit{\textbf{1(Not at all)}}} & \multicolumn{1}{l|}{\textit{\textbf{2}}} & \multicolumn{1}{l|}{\textit{\textbf{3}}} & \multicolumn{1}{l|}{\textit{\textbf{4}}} & \multicolumn{1}{l|}{\textit{\textbf{5(Expert)}}} & \multicolumn{1}{l|}{\textit{\textbf{0}}} & \multicolumn{1}{l|}{\textit{\textbf{1}}} & \multicolumn{1}{l|}{\textit{\textbf{2}}} & \multicolumn{1}{l|}{\textit{\textbf{0}}} & \multicolumn{1}{l|}{\textit{\textbf{1}}} & \textit{\textbf{2}} \\ \hline
\multicolumn{1}{|c|}{\multirow{2}{*}{<2}}   & Beginner       & 1                                                                                                                 & \multicolumn{1}{l|}{1}                               & \multicolumn{1}{l|}{0}                   & \multicolumn{1}{l|}{0}                   & \multicolumn{1}{l|}{0}                   & \multicolumn{1}{l|}{0}                           & \multicolumn{1}{l|}{1}                   & \multicolumn{1}{l|}{0}                   & \multicolumn{1}{l|}{0}                   & \multicolumn{1}{l|}{1}                   & \multicolumn{1}{l|}{0}                   & 0                   \\ \cline{2-14} 
\multicolumn{1}{|c|}{}                      & Intermediate   & 1                                                                                                                 & \multicolumn{1}{l|}{0}                               & \multicolumn{1}{l|}{1}                   & \multicolumn{1}{l|}{0}                   & \multicolumn{1}{l|}{0}                   & \multicolumn{1}{l|}{0}                           & \multicolumn{1}{l|}{1}                   & \multicolumn{1}{l|}{0}                   & \multicolumn{1}{l|}{0}                   & \multicolumn{1}{l|}{1}                   & \multicolumn{1}{l|}{0}                   & 0                   \\ \hline
\multicolumn{1}{|c|}{\multirow{2}{*}{2--5}} & Intermediate   & 3                                                                                                                 & \multicolumn{1}{l|}{0}                               & \multicolumn{1}{l|}{2}                   & \multicolumn{1}{l|}{1}                   & \multicolumn{1}{l|}{0}                   & \multicolumn{1}{l|}{0}                           & \multicolumn{1}{l|}{2}                   & \multicolumn{1}{l|}{1}                   & \multicolumn{1}{l|}{0}                   & \multicolumn{1}{l|}{3}                   & \multicolumn{1}{l|}{0}                   & 0                   \\ \cline{2-14} 
\multicolumn{1}{|c|}{}                      & Advanced       & 2                                                                                                                 & \multicolumn{1}{l|}{0}                               & \multicolumn{1}{l|}{0}                   & \multicolumn{1}{l|}{2}                   & \multicolumn{1}{l|}{0}                   & \multicolumn{1}{l|}{0}                           & \multicolumn{1}{l|}{2}                   & \multicolumn{1}{l|}{0}                   & \multicolumn{1}{l|}{0}                   & \multicolumn{1}{l|}{2}                   & \multicolumn{1}{l|}{0}                   & 0                   \\ \hline
\multicolumn{1}{|c|}{\multirow{2}{*}{>5}}   & Intermediate   & 1                                                                                                                 & \multicolumn{1}{l|}{0}                               & \multicolumn{1}{l|}{0}                   & \multicolumn{1}{l|}{1}                   & \multicolumn{1}{l|}{0}                   & \multicolumn{1}{l|}{0}                           & \multicolumn{1}{l|}{0}                   & \multicolumn{1}{l|}{1}                   & \multicolumn{1}{l|}{0}                   & \multicolumn{1}{l|}{0}                   & \multicolumn{1}{l|}{1}                   & 0                   \\ \cline{2-14} 
\multicolumn{1}{|c|}{}                      & Advanced       & 2                                                                                                                 & \multicolumn{1}{l|}{0}                               & \multicolumn{1}{l|}{0}                   & \multicolumn{1}{l|}{2}                   & \multicolumn{1}{l|}{0}                   & \multicolumn{1}{l|}{0}                           & \multicolumn{1}{l|}{1}                   & \multicolumn{1}{l|}{1}                   & \multicolumn{1}{l|}{0}                   & \multicolumn{1}{l|}{2}                   & \multicolumn{1}{l|}{0}                   & 0                   \\ \hline
\multicolumn{2}{|c|}{\textbf{Total}}                         & \textbf{10}                                                                                                       & \multicolumn{1}{l|}{\textbf{1}}                      & \multicolumn{1}{l|}{\textbf{3}}          & \multicolumn{1}{l|}{\textbf{6}}          & \multicolumn{1}{l|}{\textbf{0}}          & \multicolumn{1}{l|}{\textbf{0}}                  & \multicolumn{1}{l|}{\textbf{7}}          & \multicolumn{1}{l|}{\textbf{3}}          & \multicolumn{1}{l|}{\textbf{0}}          & \multicolumn{1}{l|}{\textbf{9}}          & \multicolumn{1}{l|}{\textbf{1}}          & \textbf{0}          \\ \hline
\end{tabular}
\label{tab:ParticipantsBackgroung}
\end{center}
\end{table*}

\emph{Participants.}
To gather user insights regarding SafEncrypt, we relied on our diverse industry network to identify and invite participants who might be interested in this study. This approach enabled us to invite participants from various backgrounds with a range of experiences. We specifically targeted developers who had previously worked with Java, without requiring prior experience with cryptographic APIs. Our aim was to gain insights from users who might be new to the topic of cryptography.

Out of 30 invitees, 10 agreed to fully participate in the study by completing the tasks and filling out the survey questionnaire. The distribution of participants, along with their experience and level of expertise in Java, as well as their knowledge of cryptography and familiarity with the Java Cryptography Architecture (JCA), is outlined in Table \ref{tab:ParticipantsBackgroung}. \\ 
According to Table \ref{tab:ParticipantsBackgroung}, 70\% of the participants had no previous experience with JCA, and 90\% were unfamiliar with Google Tink, even though they all had experience with Java.

\emph{Designing of Tasks.}
Studies have shown that cryptographic misuses are too common. On the other hand, security issues arise due to the flexibilities in Initialization Vector (IV) and key (and salt) generation, as well as the use of insecure algorithms. In SafEncrypt, we generate the IV and key for the user, and they have the opportunity to select from a static list of secure options rather than having the freedom to use an algorithm like AES-ECB. Therefore, there will not be security misuses, and we do not evaluate it from a security perspective. However, developers may still encounter issues with non-working code, encompassing the likelihood of incorrect implementation leading to exceptions or non-functional code.

We designed three symmetric encryption tasks to evaluate and compare SafEncrypt with JCA and Tink, and to check if SafEncrypt truly serves its purpose. The tasks were designed with respect to the most common encryption tasks, as investigated in Section ~\ref{sec:RealWorldScenarios}.

We assigned one of these tasks to each participant to ensure that the first task does not bias the next tasks. Each participant was assigned only one task. They were required to implement the assigned task three times using three different libraries: JCA, Tink, and SafEncrypt. There was no specific order for implementing the code; participants could choose which library to use first. The codes should be equivalent across the different libraries.

\textbf{Task 1.} Basic Symmetric Encryption:
Encrypt a random plaintext string using symmetric encryption.

\textbf{Task 2.} Password-Based Key Derivation (PBKDF2) Encryption:
Use PBKDF2 to derive an encryption key from a password. Encrypt a random plain-text string using the derived key.

\textbf{Task 3.} File-Based Encryption:
Encrypt the contents of a file using symmetric encryption.
\subsubsection{Results}
Once the committed participants completed their participation in the study, we collected the data to draw significant revelations out of it.
SafEncrypt delivers on its promises due to its design, which prevents developers from exploiting potential misuses with unregulated inputs. Upon examining the code snippets provided by participants for each task, we found that all SafEncrypt and Tink solutions were correct and secure. However, we identified security risks in 70\% of the JCA codes.

\begin{figure*}
	\centering
	\includegraphics[scale=1]
 {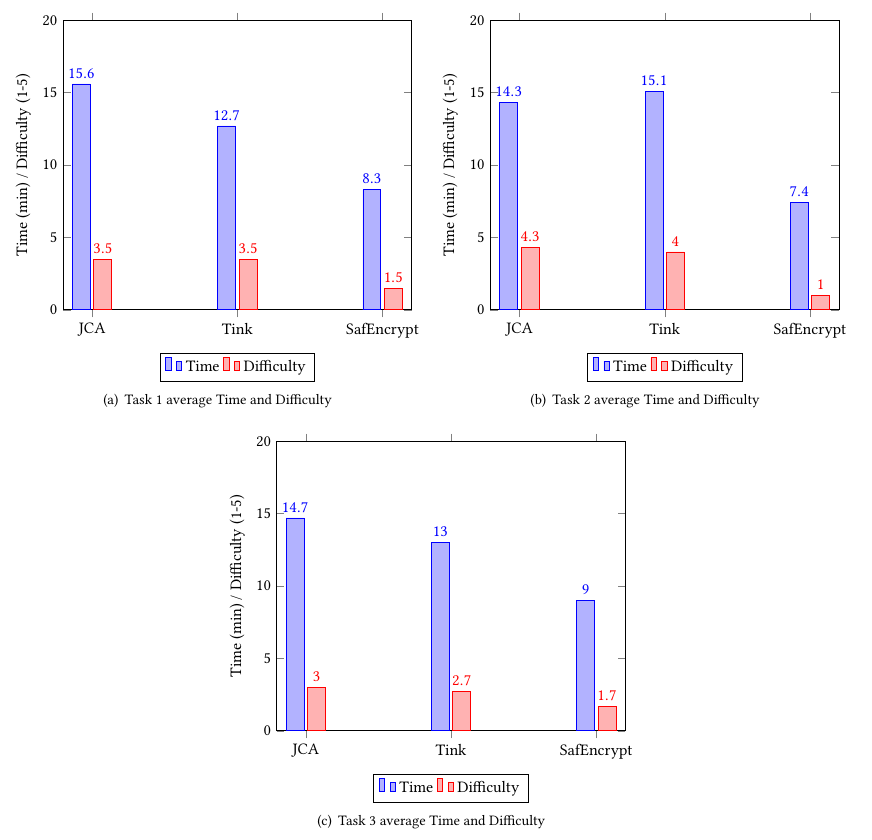}
	\caption{Average Time and Difficulty for Different Tasks}
	\label{fig:merged_tasks}
\end{figure*}

 Figure\ref{fig:merged_tasks}, illustrates the overall average time participants spent completing Tasks 1, 2, and 3, respectively, using JCA, Tink, and SafEncrypt. It also shows the difficulty ratings on a linear scale for the corresponding task.

For a detailed breakdown of the time participants spent on Tasks 1, 2, and 3, along with their difficulty ratings based on expertise, see Tables \ref{tab:Task1}, \ref{tab:Task2}, and \ref{tab:Task3}.


From Figure \ref{fig:merged_tasks} (a), which illustrates Task 1 (a simple encryption scenario), we observe that developers spent less average time (12.7 minutes) on implementation using Tink compared to JCA (15.6 minutes). However, both Tink and JCA were rated equally difficult (average difficulty rating of 3.5). SafEncrypt was found to be the easiest (average difficulty rating of 1.5) and the most time-saving option (average time of 8.3 minutes).

From Figure \ref{fig:merged_tasks} (b), which illustrates Task 2 (password-based encryption), we observe that Tink was rated slightly less difficult on average (difficulty rating of 4) than JCA (difficulty rating of 4.3). Despite this, developers found Tink more time-consuming for this task, with an average time of 15.1 minutes compared to JCA's 14.3 minutes. Additionally, Table \ref{tab:Task2} shows that beginner and intermediate developers found JCA easier for this task, while only advanced developers found Tink easier. This is likely due to Tink's lack of native support for password-based encryption. SafEncrypt was again the easiest to use (average difficulty rating of 1) and saved the most time, with all participants rating it very easy (rating of 1) and taking approximately half the time compared to JCA and Tink.

From Figure \ref{fig:merged_tasks} (c), which illustrates Task 3 (file-based encryption), we observe that JCA was the most difficult (average difficulty rating of 3) and required the most time (average time of 14.7 minutes). Tink was less difficult (average difficulty rating of 2.7) and less time-consuming (average time of 13 minutes) than JCA. SafEncrypt was again the easiest (average difficulty rating of 1.7) and the most time-efficient (average time of 9 minutes).

Overall, regardless of Java expertise, experience level, or specific task, all participants found SafEncrypt easier to understand and use compared to JCA and Tink. The results indicate that using SafEncrypt leads to significant time savings for these three common symmetric cryptography tasks.

\emph{Insights According to Java Expertise.}
The initial goal of SafEncrypt was to provide a simple and secure cryptography library for novices and developers with limited cryptography knowledge. Based on the results, we have been successful in this regard. Additionally, our findings show that it can be beneficial for developers at all levels of Java and cryptography expertise, including advanced and experienced developers.

Although JCA offers a high degree of flexibility and may be a better choice for complex scenarios, SafEncrypt is clearly the superior option for common scenarios, even for seasoned developers.

In summary, SafEncrypt has achieved its aim of providing an easy-to-use cryptography library, particularly for novice developers and those with limited cryptography knowledge.

\emph{Participants Comments and Feedback.}
Participants provided positive feedback on SafEncrypt’s design and usability. One participant highlighted the library’s advantages, noting: “This library has the potential to save developers time while offering strong implicit security features. Its highly user-friendly interface and descriptive naming conventions are noteworthy''.

Another participant mentioned an issue where they initially forgot to specify the algorithm during decryption. SafEncrypt responded with a clear exception message, which facilitated a quick resolution. They commented: “I missed specifying the algorithm as a parameter for symmetric Decryption initially, but the exception message clearly pointed out the issue and helped me resolve it. Once corrected, the process was straightforward and easy to use. Overall, the library looks promising''.

In comparison to JCA and Tink, all participants found SafEncrypt easier to use. Additionally, 90\% of them indicated that they would choose SafEncrypt for future cryptography tasks, reflecting strong user preference for the library. One participant remarked: “I faced no issues or errors with SafEncrypt, even though I have no experience with cryptography or these concepts in Java. It was an interesting and smooth experience''.

\begin{table}[]
\caption{Task 1 average Time (T) and Difficulty (D) as per Java Expertise
}
\begin{tabular}{|l|ll|ll|ll|}
\hline
\multicolumn{1}{|c|}{\multirow{2}{*}{\textbf{Java Expertise}}} & \multicolumn{2}{c|}{\textbf{JCA}}            & \multicolumn{2}{c|}{\textbf{Tink}}           & \multicolumn{2}{c|}{\textbf{SafEncrypt}}     \\ \cline{2-7} 
\multicolumn{1}{|c|}{}                                         & \multicolumn{1}{c|}{\textbf{T}} & \textbf{D} & \multicolumn{1}{l|}{\textbf{T}} & \textbf{D} & \multicolumn{1}{l|}{\textbf{T}} & \textbf{D} \\ \hline
\textbf{Beginner}                                              & \multicolumn{1}{l|}{17.2}       & 4.0        & \multicolumn{1}{l|}{12.4}       & 4.0        & \multicolumn{1}{l|}{9.3}        & 2.0        \\ \hline
\textbf{Intermediate}                                          & \multicolumn{1}{l|}{14.5}       & 3.0        & \multicolumn{1}{l|}{16.2}       & 5.0        & \multicolumn{1}{l|}{8.2}        & 1.0        \\ \hline
\textbf{Advanced}                                              & \multicolumn{1}{l|}{15.35}      & 3.5        & \multicolumn{1}{l|}{11.1}       & 3.5        & \multicolumn{1}{l|}{7.8}        & 1.5        \\ \hline

\end{tabular}
\label{tab:Task1}
\end{table}

\begin{table}[]
\caption{Task 2 average Time (T) and Difficulty (D) as per Java Expertise}
\begin{tabular}{|l|ll|ll|ll|}
\hline
\multicolumn{1}{|c|}{\multirow{2}{*}{\textbf{Java Expertise}}} & \multicolumn{2}{c|}{\textbf{JCA}}            & \multicolumn{2}{c|}{\textbf{Tink}}           & \multicolumn{2}{c|}{\textbf{SafEncrypt}}     \\ \cline{2-7} 
\multicolumn{1}{|c|}{}                                         & \multicolumn{1}{c|}{\textbf{T}} & \textbf{D} & \multicolumn{1}{c|}{\textbf{T}} & \textbf{D} & \multicolumn{1}{c|}{\textbf{T}} & \textbf{D} \\ \hline
\textbf{Beginner}                                              & \multicolumn{1}{l|}{NA}         & NA         & \multicolumn{1}{l|}{NA}         & NA         & \multicolumn{1}{l|}{NA}         & NA         \\ \hline
\textbf{Intermediate}                                          & \multicolumn{1}{l|}{15.95}      & 4.0        & \multicolumn{1}{l|}{18.7}       & 4.5        & \multicolumn{1}{l|}{8.9}        & 1.0        \\ \hline
\textbf{Advanced}                                              & \multicolumn{1}{l|}{11}         & 5.0        & \multicolumn{1}{l|}{8.0}        & 2.0        & \multicolumn{1}{l|}{4.5}        & 1.0        \\ \hline

\end{tabular}
\label{tab:Task2}
\end{table}

\begin{table}[]
\caption{Task 3 average Time (T) and Difficulty (D) as per Java Expertise}
\begin{tabular}{|l|ll|ll|ll|}
\hline
\multicolumn{1}{|c|}{\multirow{2}{*}{\textbf{Java Expertise}}} & \multicolumn{2}{c|}{\textbf{JCA}}                        & \multicolumn{2}{c|}{\textbf{Tink}}                      & \multicolumn{2}{c|}{\textbf{SafEncrypt}}                 \\ \cline{2-7} 
\multicolumn{1}{|c|}{}                                         & \multicolumn{1}{c|}{\textbf{T}} & \textbf{D} & \multicolumn{1}{l|}{\textbf{T}} & \textbf{D} & \multicolumn{1}{l|}{\textbf{Time}} & \textbf{Difficulty} \\ \hline
\textbf{Beginner}                                              & \multicolumn{1}{l|}{NA}            & NA                  & \multicolumn{1}{l|}{NA}            & NA                 & \multicolumn{1}{l|}{NA}            & NA                  \\ \hline
\textbf{Intermediate}                                          & \multicolumn{1}{l|}{14.8}          & 3.0                 & \multicolumn{1}{l|}{13.2}          & 2.5                & \multicolumn{1}{l|}{8.7}           & 1.5                 \\ \hline
\textbf{Advanced}                                              & \multicolumn{1}{l|}{14.7}          & 3.0                 & \multicolumn{1}{l|}{12.6}          & 3.0                & \multicolumn{1}{l|}{9.8}           & 2.0                 \\ \hline
\end{tabular}
\label{tab:Task3}
\end{table}

\section{Threats to Validity}
\label{sec:ThreatstoValidity}

\textbf{Threats to Internal Validity.}
Human biases and errors are inherent to manual analyses.
We implemented a meticulous review strategy. Each post underwent scrutiny by at least two individuals, and discrepancies were resolved through a majority voting approach.




Participants in this study were required to record their time. We provided clear guidelines on how to record time, including when to start and stop.
However, we cannot ensure that all participants adhered strictly to the reporting protocol.

\textbf{Threats to Construct Validity.}
The contribution of SafEncrypt might appear limited. However, we designed this API based on insights gathered from a literature review and an analysis of StackOverflow posts related to JCA and Tink. Additionally, two experts engaged in discussions to cover all key challenges, including security issues. We also evaluated whether our API provides effective solutions and ease of use for common scenarios encountered by JCA users on StackOverflow. This comprehensive approach allowed us to better assess the API’s applicability and usefulness

The experimenters' expectations could bias their interpretation of results. 
To mitigate this issue in the user study, we recruited new participants who were not involved in designing SafEncrypt.

\textbf{Threats to External Validity.}
There are threats to the generalizability of our results. 
We compared SafEncrypt with JCA and Tink, but comparison with other libraries might provide further insights.
The study included 10 participants who worked on three tasks.
We ensured that they had varying levels of experience and that the tasks were representative.
Nonetheless, further investigations in production settings with more developers will be more conclusive.
%
%

Finally, the participants worked on three tasks, and designing more tasks may offer more accurate results. To mitigate this issue, we ensured that the designed tasks were representative of the most common intended tasks mentioned on the StackOverflow website.


\section{Conclusion}
\label{sec:Conclusion}
We studied hurdles
that developers face when adopting symmetric encryption in Java
Cryptography Architecture (JCA) and investigated whether and
how they are cleared in Google Tink, which is a modern encryption
library.
We applied lessons learned from developer challenges to develop SafEncrypt, an API that is built on top of JCA. 
Through its step builder pattern it offers task-based solutions, shielding developers from
crypto complexities and erroneous low-level details, and making encryption process intuitive for Java developers, particularly those new to cryptography.

This work showed the feasibility of building a cryptography API that is easy to use for developers and serves common cryptography scenarios in Java programs.

SafEncrypt currently supports symmetric and streaming symmetric encryption operations, and incorporation of additional operations such as asymmetric encryption remains a future plan

\bibliographystyle{ACM-Reference-Format}
\bibliography{paper}

\end{document}